\documentclass[preprint,showpacs,preprintnumbers,amsmath,amssymb]{revtex4}

\usepackage{graphicx}
\usepackage{dcolumn}
\usepackage{bm}

\begin{document}

\preprint{submitted to Physical Review Letters}

\title{Formation and light guiding properties of dark solitons in one-dimensional waveguide arrays}

\author{Eugene Smirnov}
\author{Christian E. R\"uter}
\author{Milutin Stepi\'c}
\author{Detlef Kip}
\email{d.kip@pe.tu-clausthal.de}
\affiliation{Institute of Physics
and Physical Technology, Clausthal University of Technology, 38678
Clausthal-Zellerfeld, Germany}

\author{Vladimir Shandarov}
\affiliation{State University of Control Systems and
Radioelectronics, 40 Lenin Ave., 634050 Tomsk, Russia}

\date{\today}


\begin{abstract}
We report on the formation of dark discrete solitons in a
nonlinear periodic system consisting of evanescently-coupled
channel waveguides that are fabricated in defocusing lithium
niobate. Localized nonlinear dark modes displaying a phase jump in
the center that is located either on-channel (mode A) or
in-between channels (mode B) are formed, which is to our knowledge
the first experimental observation of mode B. By numerical
simulations we find that the saturable nature of the nonlinearity
is responsible for the improved stability of mode B. The ability
of the induced refractive index structures to guide light of a
low-power probe beam is demonstrated.
\end{abstract}

\pacs{42.65tg, 42.65Wi, 42.82Et, 63.20Pw}

\maketitle


Nonlinear wave propagation in periodic lattices, which occurs in
many different systems in nature \cite{A1,A2,1,A3}, has attracted
great interest in recent years. In these systems the dynamics is
dominated by an interplay of diffraction, i.e.\ tunnelling through
adjacent potential wells, and nonlinearity, leading to a large
variety of nonlinear effects that have no analogue in bulk media.
This increasing interest may be at least partially attributed to
recent progress in the investigation of nonlinear wave propagation
in optical periodic media, where the ability to engineer optical
band structures and diffraction as well as a rather easy
experimental observation of nonlinear effects has led to the
discovery of many new fundamental features \cite{2,3,4,5a,5b,5c}.

In periodic optical systems including evanescently-coupled
waveguide arrays \cite{6,7}, photonic lattices and crystals
\cite{8}, different types of localized bright structures or
lattice solitons have been observed. To name a few, theoretical
prediction and experimental realization of spatial gap solitons
\cite{9a,9b,10}, solitons in higher-bands \cite{11}, and solitons
occupying modes of several bands \cite{12} has been performed.
Here we again want to note that the study of these intrinsically
localized modes is a universal problem and relevant to many
non-optical systems, like localized voltage drops in ladders of
Josephson junctions \cite{X1}, localized modes in
antiferromagnetic crystals \cite{X2}, or localization of matter
waves in Bose--Einstein condensates using optically-induced
periodic potentials \cite{X3}.

For bright discrete solitons in Kerr media it has been shown
theoretically that, for a given power, two stationary localized
modes may exist \cite{E1}: a stable mode A centered on a waveguide
and an unstable mode B centered between two neighboring
waveguides. On the other hand, we have shown recently that a
saturable nonlinearity, for example by using photorefractive
nonlinear media, may also support stable propagation of mode B
\cite{E2,E3}. Very recently, similar predictions have been made
for other nonlinear media, for example in cubic-quintic systems
\cite{E4} and for localized surface waves \cite{E5}.

As is now well understood, bright lattice solitons can exist
either in the region of normal diffraction as a result of a
self-focusing (positive) nonlinearity, or in media exhibiting a
self-defocusing (negative) nonlinear index change and anomalous
diffraction of light in the lattice. Similar to the situation in
the bulk \cite{13,14}, for dark solitons normal diffraction of a
narrow dark notch (a small number of dark elements) on an
otherwise homogeneously excited lattice can be balanced by a
negative index change, or alternatively anomalous diffraction in a
lattice can be compensated by a positive nonlinearity. In periodic
systems, the existence of dark discrete solitons has been
investigated theoretically \cite{15}, followed only recently by
first experimental realizations \cite{10,16}. As has been
recognized already in bulk materials, dark solitons are potential
candidates for guiding, steering, and switching of light beams in
light-induced refractive index channels. Even more interesting, in
discrete media like coupled waveguide arrays, a realization of the
above mentioned functions would strongly benefit from the inherent
multi-port structure of the array.

In this Letter we investigate formation of dark discrete solitons
in one-dimensional waveguide arrays in a material exhibiting a
saturable self-defocusing nonlinearity. Localized nonlinear dark
modes displaying a phase jump that is located either on (mode A)
or in-between (mode B) them are formed, which is --- to our
knowledge --- the first experimental observation of mode B.
Numerical simulations that support our experimental findings show
that the saturable nature of the used nonlinearity increases
stability of mode B when compared to the purely Kerr-type
defocusing case. Furthermore, to demonstrate the ability of both
types of dark discrete solitons for light steering and switching,
guiding of probe beams that are launched into the light-induced
refractive index structures is demonstrated.


The waveguide array is fabricated in a Cu-doped lithium niobate
crystal, where the saturable defocusing optical nonlinearity
arises from the bulk photovoltaic effect. The array that is
fabricated by in-diffusion of titanium consists of approximately
250 channels with a width of 4.4\,$\mu$m and a grating period of
8.4\,$\mu$m \cite{9b,17}.

\begin{figure}[t]
\centerline{\includegraphics[width=8.5cm]{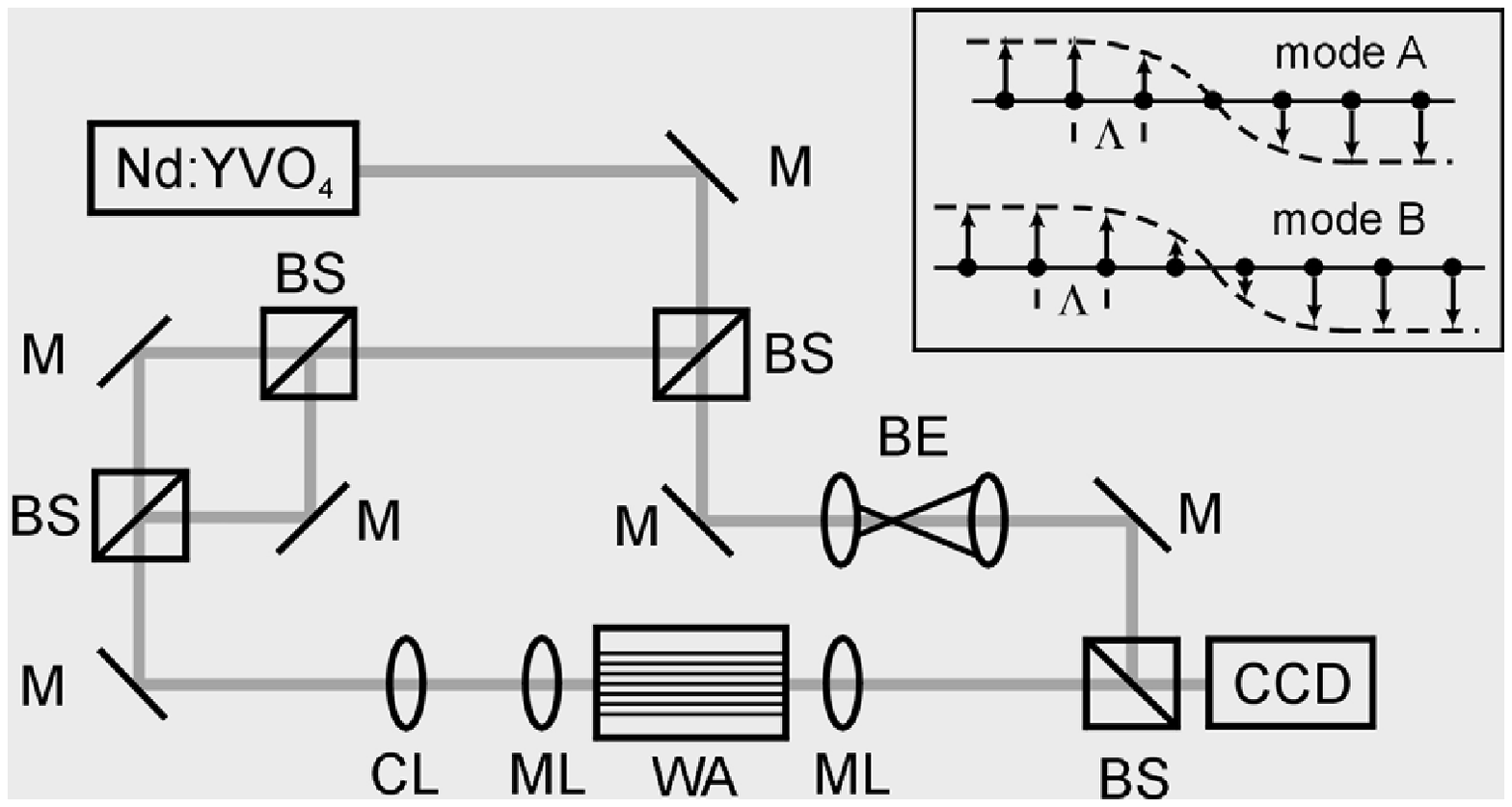}}
\caption{Experimental setup. M's, mirrors; BS's, beam splitters;
CL, cylindrical lens; ML's, microscope lenses; WA, waveguide
array; BE, beam expander; CCD, CCD camera. Inset: phase profiles
of modes A and B.}
\end{figure}

The experimental setup is sketched in Fig.~1. We use a
frequency-doubled Nd:YVO$_{4}$ laser with a wavelength $\lambda =
532$\,nm. The light is split into three beams, where two of them
are formed by a Michelson interferometer and partially
superimposed under a small angle on the input face of the
waveguide array. In this way a broad beam covering about 25
channels with a small dark notch caused by destructive
interference in the overlap region is formed. The center of the
formed input beam that experiences a phase jump of $\pi$ in its
center can be adjusted either on-channel to excite mode A, or
in-between channels to excite mode B (see inset of Fig.~1). A
third beam that is expanded to a plane wave with the help of a
beam expander is used to investigate the phase structure of the
guided light. For this the plane wave interferes with the
outcoupled light of the array on the CCD camera.

\begin{figure}[t]
\centerline{\includegraphics[width=8.5cm]{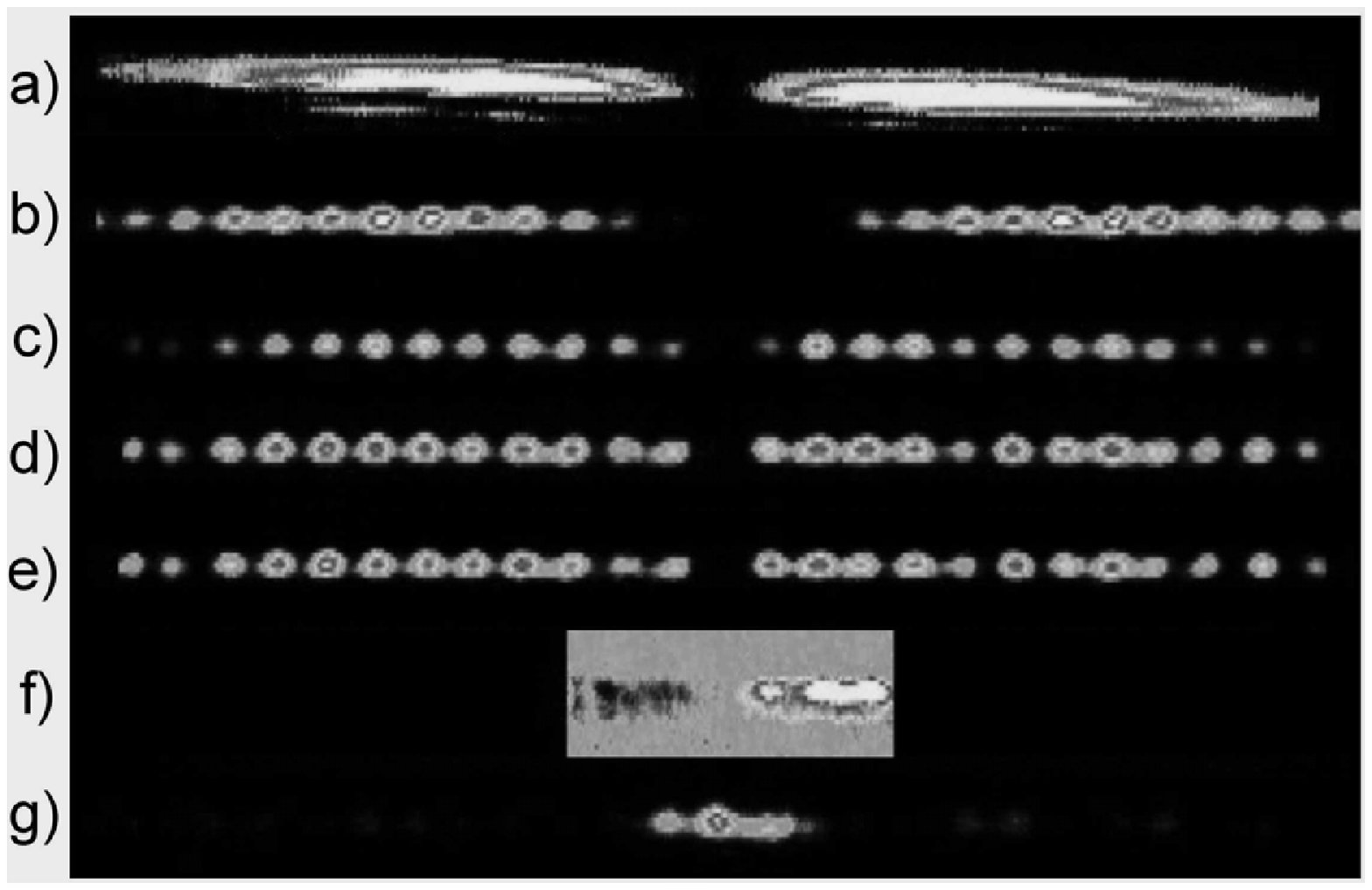}}
\caption{Discrete diffraction and nonlinear light propagation of
mode A: a) intensity distribution on the sample's input face, b)
discrete diffraction of mode A ($P_{in}=1\,\mu$W), c) nonlinear
partial focusing at $t=10$\,sec, and d) and e) dark soliton
formation at $t=20$\,sec and $t=120$\,s, respectively
($P_{in}=100\,\mu$W), f) interferogram of the dark soliton shown
in part d), and g) guiding of a weak probe beam coupled into the
central channel.}
\end{figure}

In a first experiment we investigate the linear and nonlinear
propagation of mode A, where the phase discontinuity is located on
a lattice element, i.e.\ the central channel is hardly excited.
Here the input width (FWHM) of the dark notch with a total optical
power of 100\,$\mu$W is about $\Delta z=25\,\mu$m, covering
roughly 4 periods of the lattice and propagating in forward
direction with the transverse wave vector component equal to zero.
To monitor the input intensity, the photograph in Fig.~2a) shows
the reflected light from the sample's input face. In the linear
case for low input power, discrete diffraction in Fig.~2b) leads
to a broadening of the structure on the homogeneous background,
which reaches about 5 channels on the output face after 11\,mm of
propagation. The response time of the photovoltaic nonlinearity in
our sample is about 20\,sec for the used input power, thus we are
able to monitor the build-up of the discrete dark soliton as a
function of time. When the power is increased, first nonlinear
de-focusing starts (Fig.~2c) and eventually forms a dark soliton
in Fig.~2d) after about 20\,sec. This structure is stable over
times large compared to the build-up time of the nonlinearity
(Fig.~2e), taken after $t=120\,$sec).

To investigate the nonlinearly induced refractive index structure,
we measure the phase profile of the dark soliton by interfering it
with a plane wave in Fig.~2f). Obviously, the resulting trapped
state has conserved the input phase discontinuity from the input.
When the input light is switched off, a probe beam can be coupled
into the central input channel. As can be seen in Fig.~2g), the
induced structure forms a single mode waveguide guiding the light
of this probe beam.

\begin{figure}[b]
\centerline{\includegraphics[width=8.5cm]{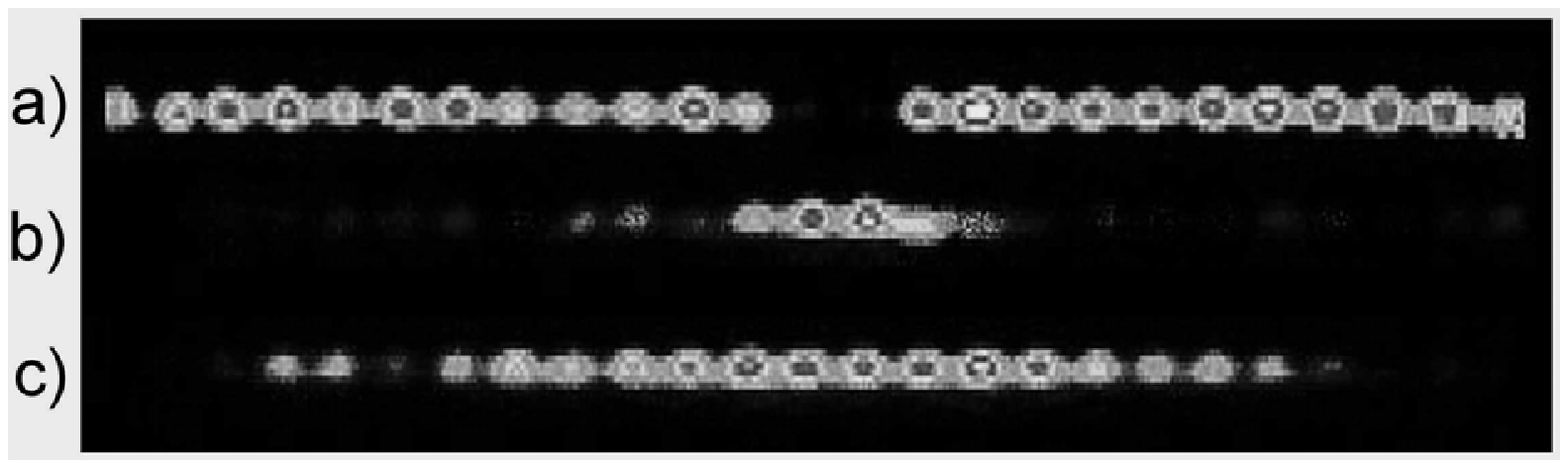}}
\caption{Nonlinear propagation of mode B: a) dark soliton
formation at $t=60$\,sec, b) guiding of a probe beam, and c)
diffraction of the input probe beam in the linear lattice.}
\end{figure}

In a second experiment, the waveguide array is laterally shifted
by half a lattice period with respect to the interfering input
beams, thus exciting mode B with a phase jump in-between two
channels. For a very similar input power as for mode A, an even
dark discrete soliton is formed with two dark channels in the
center in Fig.~3a). This is to our knowledge the first
experimental observation of this intrinsically localized mode.
Again, after switching off the pump light the induced structure
can be probed by another beam, which in this case covers about 4
channels on the input face. Obviously, this probe beam is guided
in two parallel waveguide channels displayed in Fig.~3b). When
illuminating the sample with intense white light, the induced
refractive index structure is erased and finally the probe beam
propagates with normal diffraction in the (undisturbed) lattice in
Fig.~3c).

\begin{figure}[t]
\centerline{\includegraphics[width=8.5cm]{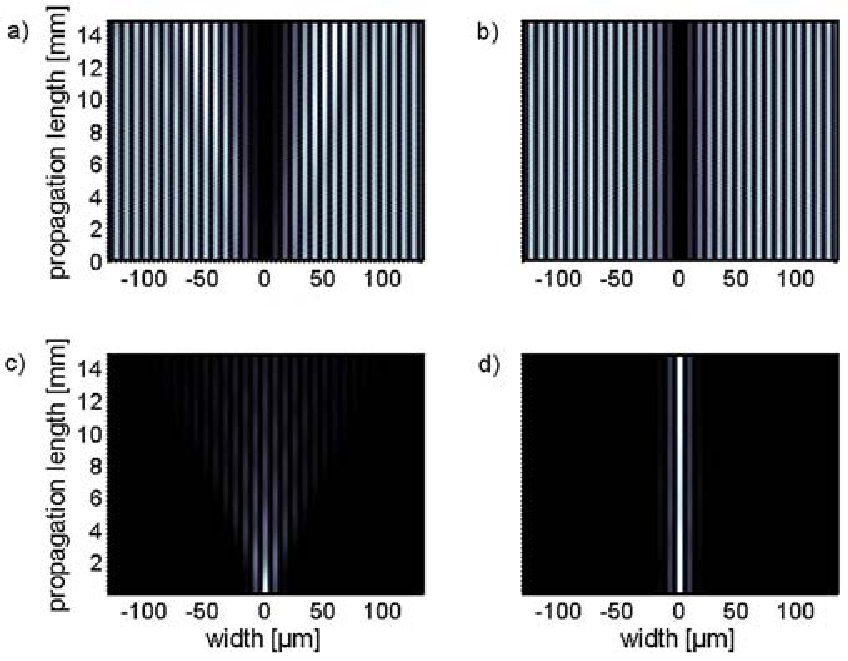}}
\caption{Simulation of mode A: a) linear diffraction of a dark
notch, b) discrete dark soliton formation, c) diffraction and d)
guiding of a probe beam, respectively.}
\end{figure}

To compare our experimental results with theory, we simulate the
light propagation in the array using a nonlinear beam propagation
method. For mode A with an input width (FWHM) of $\Delta
z=25\,\mu$m, in Fig.~4a) the case of linear diffraction for low
input power is shown. A localized dark soliton is obtained in
Fig.~4b) in the nonlinear regime applying a saturable nonlinearity
of the form $\Delta n=\Delta n_{0}\,r/(1+r)$, where $\Delta
n_{0}=-1.5\times 10^{-4}$ is the amplitude of nonlinear index,
$r=I/I_{d}=8$ is light intensity ratio with intensity $I$ and dark
irradiance $I_{d}$. When the dark soliton is formed, a low-power
probe beam that is coupled into the central channel can be guided
in the written refractive index distribution of the lattice
(Fig.~4d). In the linear case this beam diffracts without guiding
(Fig.~4c).

\begin{figure}[t]
\centerline{\includegraphics[width=8.5cm]{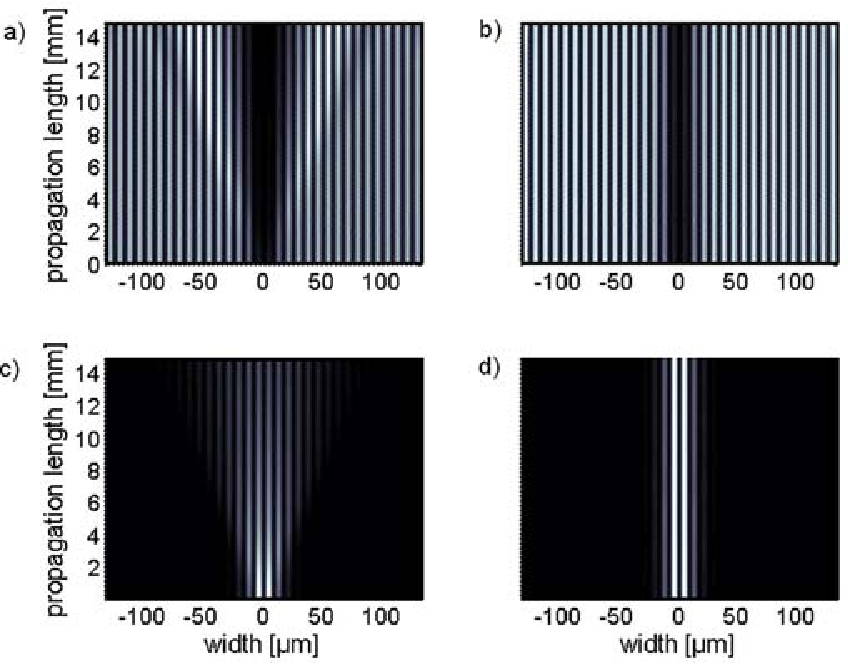}}
\caption{Simulation of mode B: a) discrete diffraction of a linear
beam, b) discrete dark soliton formation, c) diffraction and d)
guiding of a probe beam, respectively.}
\end{figure}

For mode B and the same nonlinearity as above, linear diffraction
and soliton formation are shown in Fig.~5a) and b), in good
agreement with the experimental results. For a broader (linear)
input probe beam covering about four input channels, the resulting
even refractive index distribution shows guiding in the two
channels (Fig.~5d)), whereas this beam diffracts to a broad output
beam in the linear case.

As has been shown theoretically for the Kerr case \cite{15}, mode
B experiences instability during propagation. Here instability
(i.e., conversion of mode B into mode A) may originate from a
small mismatch of the symmetric input condition, for example by a
lateral shift of the input light distribution, a small tilt angle
of the input beam or a small asymmetry in intensity. However, our
simulation results show that the stability of mode B is
significantly improved by the saturable form of the nonlinearity,
as it has been predicted recently in Ref.~\cite{E2} for bright
discrete solitons.

\begin{figure}[b]
\centerline{\includegraphics[width=8.5cm]{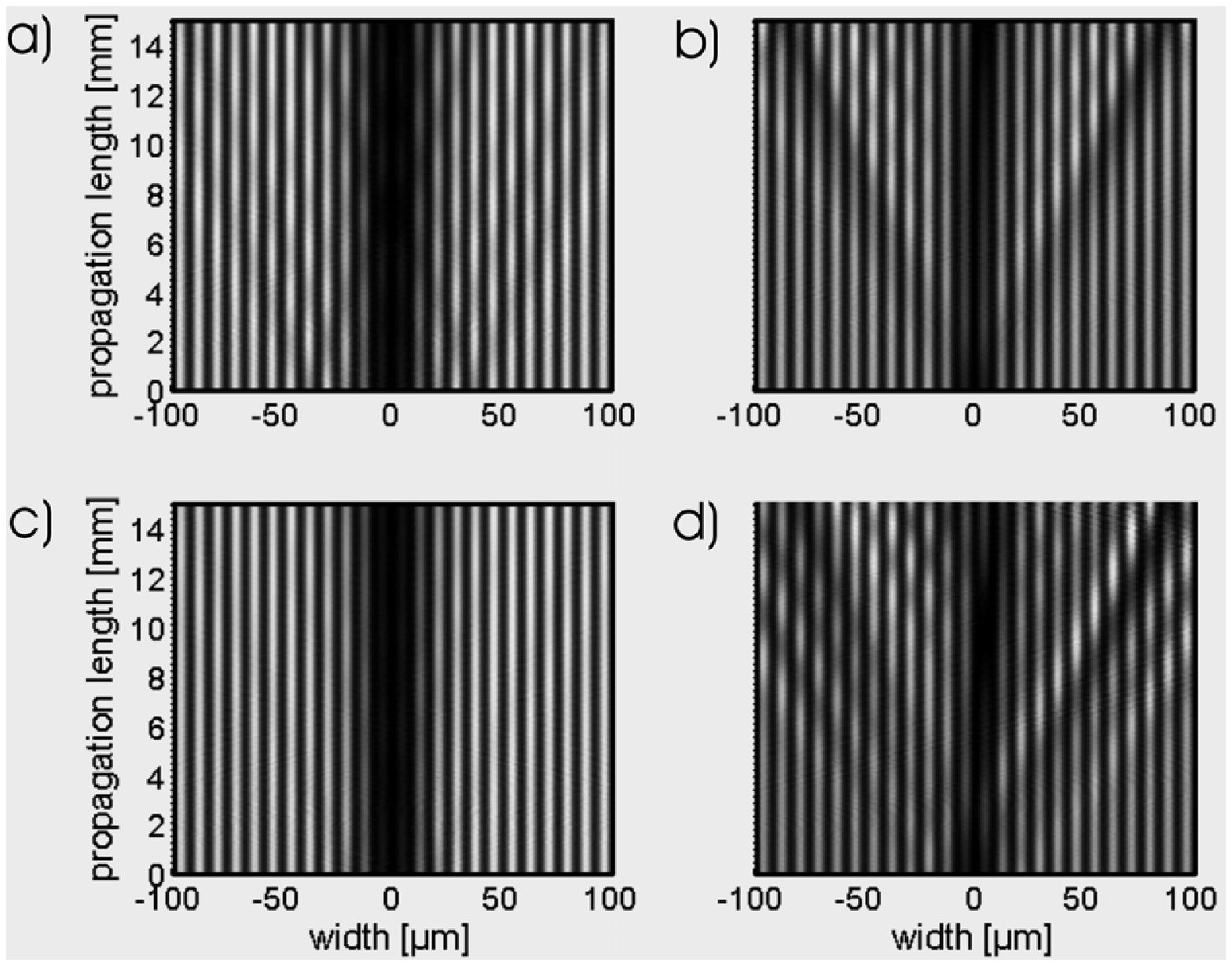}}
\caption{Comparison of stability of mode B for saturable and
Kerr-type nonlinearity. Tilt angle $\delta\alpha=0.5\,^{\circ}$ of
the input phase front for a) saturable and b) Kerr case, and
asymmetry $P_{in,r}/P_{in,l}=1.03$ of power of right and left half
of the input beam for c) saturable and d) Kerr case.}
\end{figure}

Two numerical examples of the improved stability are given in
Fig.~6. In the upper part (Fig.~6a) and b)) we have added a small
tilt angle $\delta\alpha=0.5\,^{\circ}$ of the input phase front,
with all other parameters being the same as in Fig.~5. As can be
seen, for a saturable nonlinearity (a) almost stable propagation
of mode B is  obtained, while for a purely Kerr-type nonlinearity
the input mode experiences instability and is converted into mode
A.

A similar behavior is observed if we introduce a small asymmetry
in the input light intensity. In Fig.~6c) and d) the right hand
side of the input intensity is increased by 3\,\% relative to that
on the left hand side. Obviously, this asymmetry leads to
destabilization of mode B in the Kerr case in d), while the same
input intensity propagates stable in a saturable nonlinear
waveguide array in part c).

To summarize, we have experimentally investigated dark soliton
formation in a nonlinear waveguide array with defocusing saturable
nonlinearity. Stable nonlinear light propagation of localized dark
beams centered either on-site or in-between two channels has been
observed experimentally, and our findings are well supported by
numerical simulations. It has been shown that the saturable
character of the nonlinearity leads to stabilization of mode B.
The ability of the induced refractive index patterns to guide
light in form of two different stable structures has been
demonstrated, which is of great practical interest for the
realization of all-optical devices like routers and switches.

\begin{acknowledgments}
We gratefully acknowledge financial support from the German
Federal Ministry of Education and Research (BMBF grant DIP-E6.1)
and the Deutsche Forschungsgemeinschaft (DFG grants KI482/8-1 and
436RUS17/26/06). D.~Kip's e-mail address is:
d.kip@pe.tu-clausthal.de.
\end{acknowledgments}

\end{document}